\documentclass[twocolumn,showpacs,preprintnumbers,amsmath,amssymb]{revtex4}
\usepackage{graphicx}
\usepackage{epsfig}
\usepackage{dcolumn}
\usepackage{bm}

\newcommand{\ccb}{c\overline{c}}

\newcommand{\psip}{\psi(2S)}
\newcommand{\psp}{\psi(2S)}
\newcommand{\jpsi}{J/\psi}
\newcommand{\rar}{\rightarrow}
\newcommand{\ra}{\rightarrow}

\newcommand{\pipi}{\pi^+ \pi^- }
\newcommand{\piz}{\pi^0}

\newcommand{\ks}{K^0_S}
\newcommand{\xcj}{\chi_{cJ}}
\newcommand{\etap}{\eta^{\prime}(958)}
\newcommand{\etaf}{\eta(1405)/\eta(1475)}
\newcommand{\etafl}{\eta(1405)}
\newcommand{\etafh}{\eta(1475)}
\newcommand{\xc}{\chi_{c1}}

\newcommand{\api}{a_0^{\pm}\pi^{\mp}}
\newcommand{\ft}{f_2(1270)}
\newcommand{\aon}{a_0(980)}

\newcommand{\psipto}{\psi(2S)\rightarrow}
\newcommand{\pspto}{\psi(2S)\rightarrow}
\newcommand{\jpsito}{J/\psi\rightarrow}
\newcommand{\gamg}{\gamma\gamma}

\newcommand{\GeV}{\hbox{~GeV}}

\newcommand{\gev}{\,\mbox{GeV}}
\newcommand{\mev}{\,\mbox{MeV}}

\newcommand{\kk}{K^+K^-}

\newcommand{\kskp}{K^0_S K^+ \pi^- + c.c.}

\newcommand{\gkkp}{\gamma K \overline{K} \pi}

\newcommand{\chicJ}{{\chi_{cJ}}}
\newcommand{\chicj}{\chi_{cJ}}
\newcommand{\chicz}{{\rm\chi_{c0}}}
\newcommand{\chico}{{\rm\chi_{c1}}}
\newcommand{\chict}{{\rm\chi_{c2}}}

\newcommand{\BR}{{\cal B}}

\newcommand{\eff}{\scalebox{1.0}{$\varepsilon$}}

\newcommand{\kbar}{\overline{K}}
\newcommand{\kzbar}{\overline{K}^0}

\def\Journal#1&#2&#3(#4){#1{\bf~#2}, #3 (#4)}

\def\PLB{Phys.  Lett.   B}
\def\PRL{Phys.  Rev.  Lett.  }
\def\PRD{Phys.  Rev.   D}

\def\etal{{\it et al.}}

\def\bec{\begin{center}}
\def\eec{\end{center}}

\begin{document}
\title{\boldmath Measurements of $\psip$ decays into
$\gamma K\overline{K}\pi$ and $\gamma\eta\pipi$  }
\author{
M.~Ablikim$^{1}$,         J.~Z.~Bai$^{1}$,               Y.~Ban$^{12}$,
J.~G.~Bian$^{1}$,         X.~Cai$^{1}$,                  H.~F.~Chen$^{17}$,
H.~S.~Chen$^{1}$,         H.~X.~Chen$^{1}$,              J.~C.~Chen$^{1}$,
Jin~Chen$^{1}$
% ...
\\
\vspace{0.2cm}
(BES Collaboration)\\
\vspace{0.2cm}
{\it
$^{1}$ Institute of High Energy Physics, Beijing 100049, People's Republic of China\\
% ...
$^{12}$ Peking University, Beijing 100871, People's Republic of China\\
% ...
$^{17}$ University of Science and Technology of China, Hefei 230026, People's Republic of China\\
% ...
}
} 

\begin{abstract}
  Radiative decays of the $\psip$ into $\gamma K\overline{K}\pi$ and 
  $\gamma\eta\pipi$ final states are studied using 14 million 
  $\psip$ events collected with the BESII detector. Branching 
  fractions or upper limits on the branching fractions of 
  $\psip$ and $\xcj$ decays are reported.
  No significant signal  for $\etafl/\etafh$ is observed in
  the $K\overline{K}\pi$ or $\eta\pipi$ mass spectra, 
  and upper limits on the branching 
  fractions of $\psip\rar\gamma\etaf$,   $\etaf\rar K\overline{K}\pi$ 
  and $\eta\pipi$ are determined.
\end{abstract}
\maketitle

\section{\boldmath Introduction}
$\psip$ decays via three gluons or a single direct photon have been
extensively studied~\cite{pQCDrule}. However, there have been
fewer studies of $\psip$ radiative decays~\cite{PDG04}. Further study of
$\psip$ radiative decays will provide more information about the
$\psip$ decay mechanism and may help in understanding problems like
the ``$\rho\pi$ puzzle''. The ``12\% rule'' predicted by perturbative
QCD~\cite{qcd15} is expected to be applicable to $\psip$ radiative
decays~\cite{Jdecay}, so it can be tested by measuring more of these
decays.  Furthermore, if the 12\% rule is obeyed for the
$\psip\rar\gamma\eta(1440)$~\cite{PDG04} decay, we might expect to
observe $\eta(1440)$ in $\psip$ decays into $\gamma K\overline{K}\pi$
and $\gamma\eta\pipi$.

A glueball candidate,  the $\eta(1440)$, is  now
regarded as the superposition
of two independent states, the $\eta(1405)$ and the $\eta(1475)$, with
different decay modes~\cite{PDG04}. 
The $\eta(1475)$ could be the first radial excitation of the $\etap$,
while the $\eta(1295)$ could be the first radial excitation of the
$\eta$.  The results of L3's measurements on the $K\overline{K}\pi$
and $\eta\pipi$ channels in $\gamma\gamma$ collisions suggest that the
$\eta(1405)$ has a large gluonic content~\cite{L32001}.  However CLEO
did not confirm L3's results with a five times larger data sample and
set upper limits on
$\Gamma_{\gamg}(\eta(1405))\mathcal{B}(\eta(1405)\rar
K\overline{K}\pi)$ and
$\Gamma_{\gamg}(\eta(1475))\mathcal{B}(\eta(1475)\rar
K\overline{K}\pi)$, which are still consistent with the glueball and
the radial excitation hypotheses for $\eta(1405)$ and
$\eta(1475)$~\cite{prd072001}.

Many studies have been made for $\eta(1405)/\eta(1475)$ with $J/\psi$
decays into $K\overline{K}\pi$, $\eta\pipi$, $4\pi$, and
$\gamma\rho^0$~\cite{PDG04}, while in $\psip$ decay, only MARKI
reported an upper limit at the 90\% confidence level (C.L.) for
$\psipto\gamma \eta(1405)\rar\gamma K\overline{K}\pi$~\cite{markI80}.
Here we study $\eta(1405)/\eta(1475)$ in $\psip$ radiative decays to
$\gamma K\overline{K}\pi$ and $\gamma\eta\pipi$ final states using a
sample of $14\times10^6$ $\psip$ events.

In lowest-order perturbative QCD, the $\chicz$ and $\chict$ 
decay via the annihilation of their constituent $\ccb$ quarks 
into two gluons, followed by hadronization of the gluons
into light mesons and baryons, so these decays are expected to
be similar to those of a $gg$ bound state, while $\chi_{c1}$
cannot decay via the annihilation of their constituent $\ccb$ quarks 
into two gluons. So systematic and  detailed studies
of hadronic decays of the $\chi_{cJ}$ may help
in understanding the decay patterns of glueball states 
that will be helpful for their identification.

BESI collaboration studied $\chi_{cJ}$ decays into 
$\kskp$~\cite{besichic} and reported $\chi_{c1}$ branching fractions 
and upper limits on branching fractions of
 $\chi_{c0}$ and $\chi_{c2}$ decays.
In this paper, we report measurements of $\psi(2S)$ decays into  
$\gamma K\overline{K}\pi$ and $\gamma\eta\pipi$ final states using 
14 million $\psip$ events collected with the BESII 
detector. With about a four times larger $\psip$ sample, more 
precise results are expected.
Branching fractions or upper limits 
of $\psip$ and $\xcj$ decays are reported.

\section{\boldmath THE BESII DETECTOR}
The Beijing Spectrometer (BESII) is a conventional cylindrical
magnetic detector that is described in detail in Ref.~\cite{BES-II}.
A 12-layer vertex chamber (VC) surrounding the beryllium beam pipe
provides input to the event trigger, as well as coordinate
information.  A forty-layer main drift chamber (MDC) located just
outside the VC yields precise measurements of charged particle
trajectories with a solid angle coverage of $85\%$ of $4\pi$; it also
provides ionization energy loss ($dE/dx$) measurements which are used
for particle identification.  Momentum resolution of
$1.7\%\sqrt{1+p^2}$ ($p$ in GeV/$c$) and $dE/dx$ resolution for hadron
tracks of $\sim8\%$ are obtained.  An array of 48 scintillation
counters surrounding the MDC measures the time of flight (TOF) of
charged particles with a resolution of about 200 ps for hadrons.
Outside the TOF counters, a 12 radiation length, lead-gas barrel
shower counter (BSC), operating in limited streamer mode, measures the
energies of electrons and photons over $80\%$ of the total solid angle
with an energy resolution of $\sigma_E/E=0.22/\sqrt{E}$ ($E$ in GeV).
A solenoidal magnet outside the BSC provides a 0.4 T magnetic field in
the central tracking region of the detector. Three double-layer muon
counters instrument the magnet flux return and serve to identify muons
with momentum greater than 500 MeV/$c$. They cover $68\%$ of the total
solid angle.

\section{\boldmath Event Selection}
The decay channels
investigated in this paper are $\psipto\gamma \ks K^+\pi^-+c.c.$, 
$\gamma K^+K^-\pi^0$ and $\gamma\eta\pipi$,  
where $\ks$ decays to $\pi^+\pi^-$, $\eta$ to $\gamma\gamma$,  
and $\pi^0$ to $\gamma\gamma$. 

A neutral cluster is considered to be a photon candidate if the
following requirements are satisfied: it is located within the BSC
fiducial region, the energy deposited in the BSC is greater than 50 MeV, 
the first hit appears in the first 6 radiation lengths, the angle in the
$x y$ plane (perpendicular to the beam direction) between the cluster
and the nearest charged track is greater than $8^\circ$, and the angle 
between the cluster development direction in
the BSC and the photon emission direction from the beam interaction
point (IP) is less than $37^\circ$.

Each charged track is required to be well fitted by a
three-dimensional helix, to have a momentum transverse to the beam
direction greater than 70 MeV/$c$, to originate from the IP region
($V_{xy}=\sqrt{V_x^2+V_y^2}<2$ cm and $|V_z|<20$ cm) if it is not 
from $\ks$ decay, and to have a polar angle $|\cos\theta|<0.8$. 
Here $V_x$, $V_y$, and $V_z$ are the
$x$, $y$, and $z$ coordinates of the point of closest approach of the
track to the beam axis.

The TOF and $dE/dx$ measurements for each charged track are used to
calculate $\chi^2_{PID}(i)$ values and the corresponding confidence 
levels $Prob_{PID}(i)$ for the hypotheses that a track is a pion, 
kaon, or proton, where $i$ ($i=\pi/K/p$) is the particle type.
For each event,  charged kaon candidates are required to have 
$Prob_{PID}(K)$ larger than 0.01, while charged pion candidates
are required to have $Prob_{PID}(\pi) >0.01$.

\section{\boldmath Event analysis}

\subsection{\boldmath $\psip\rar\gamma \ks K^+\pi^-+c.c.$}
For the final state $\gamma K^\pm\pi^\mp\pi^+\pi^-$, the
candidate events are required to have at least one photon candidate and 
four good charged tracks with net charge zero.
A four constraint (4C) kinematic fit is performed to the hypothesis
$\psip\rar\gamma K^\pm\pi^\mp\pi^+\pi^-$, and the $\chi^2$ of the fit is
required to be less than 15. If there is more than one photon,
the fit is performed with the photon candidate which has the largest energy
deposit in the BSC. A 4C-fit to the hypothesis
$\psip\rar\gamma \pi^+\pi^-\pi^+\pi^-$ is also performed, and
$\chi^{2}_{4C}(\gamma K^\pm\pi^\mp\pi^+\pi^-)<
\chi^{2}_{4C}(\gamma \pi^+\pi^-\pi^+\pi^-)$ is required to suppress
background from $\gamma \pi^+\pi^-\pi^+\pi^-$.
\begin{figure}[htb]
\centerline{\hbox{\psfig{file=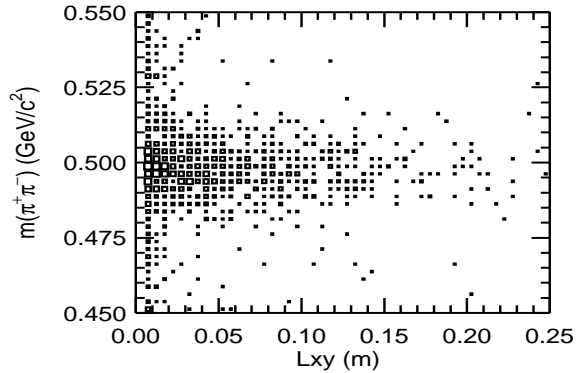,width=0.45\textwidth ,height=5cm}}}
\caption{The scatter plot of $\pi^+\pi^-$ invariant mass versus the $\ks$ decay length.}
\label{mks-lxy}
\end{figure}

Backgrounds from $\psip\rar\pipi J/\psi$ are rejected with the
requirement $|m^{\pi^{+}\pi^{-}}_{recoil}-3.1|>0.05$ GeV/$c^2$, where
$m^{\pi^{+}\pi^{-}}_{recoil}$ is the mass recoiling from each possible
$\pipi$ pair.  Figure~\ref{mks-lxy} shows the scatter plot of
$\pi^+\pi^-$ invariant mass versus the decay length in the transverse
plane ($L_{xy}$) of $\ks$ candidates. A clear $\ks$ signal is
observed.  Candidate events are required to have only one $\ks$
candidate satisfying the requirements $|m_{\pipi}-0.498|<0.015$
GeV/$c^2$ and $L_{xy}>0.5$ cm.  After $\ks$ selection, if one of the
remaining tracks has a momentum higher than 1.5 GeV/$c$, it is taken
as a charged kaon.  Otherwise, the track types are selected 
using their
$\chi^2_{K\pi}$ values, \textit{i.e.}, if
$\chi^2_{K^+\pi^-}<\chi^2_{\pi^+K^-}$, the final state is considered
to be $\gamma \ks K^+\pi^-$; if $\chi^2_{K^-\pi^+}<\chi^2_{\pi^-K^+}$,
the final state is considered to be $\gamma \ks K^-\pi^+$, where
$\chi^2_{K\pi}=\chi^2_{PID}(K)+\chi^2_{PID}(\pi)$.

\begin{figure}[htb]
\centerline{\hbox{\psfig{file=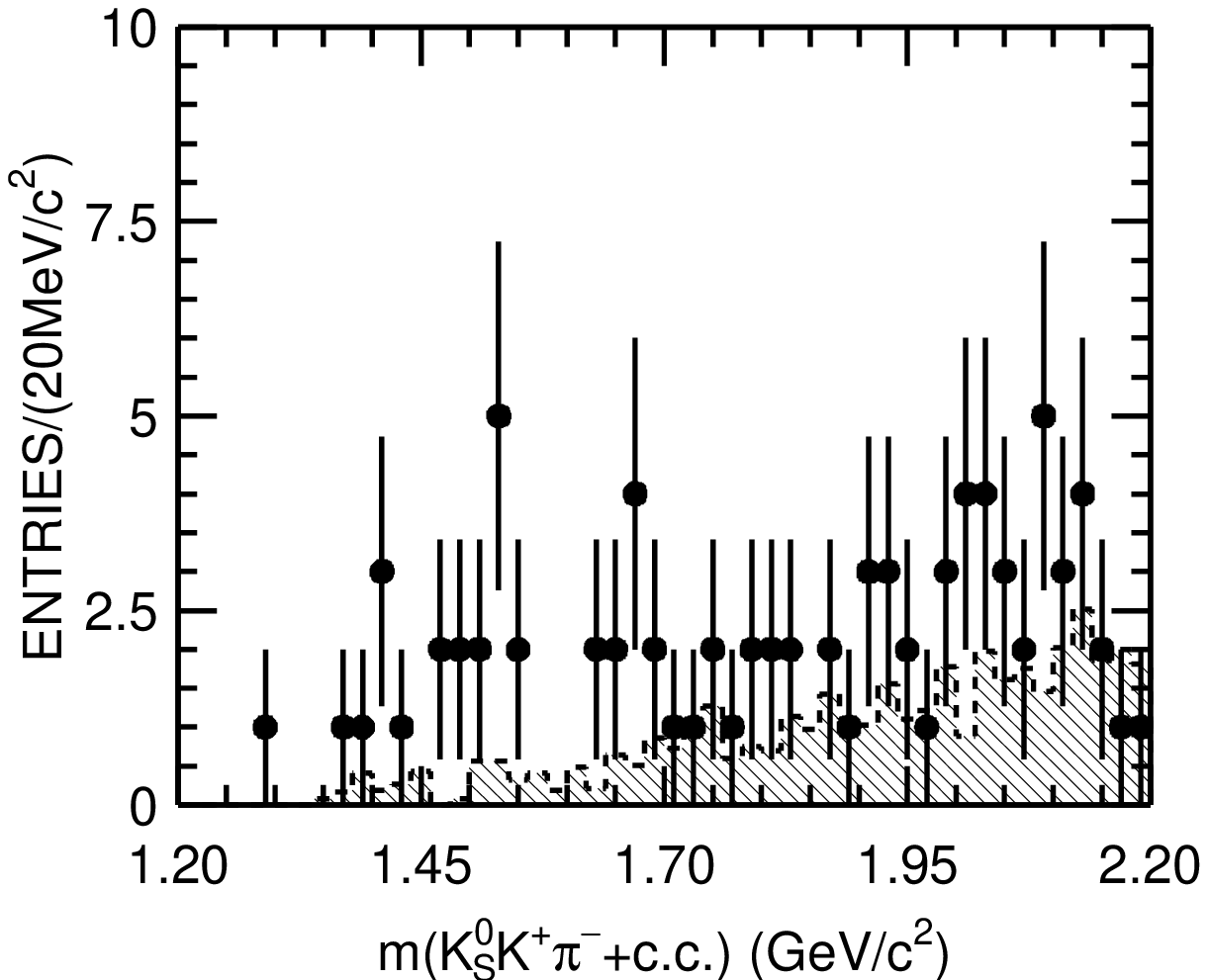,width=0.45\textwidth,height=5.0cm}}}
\centerline{\hbox{\psfig{file=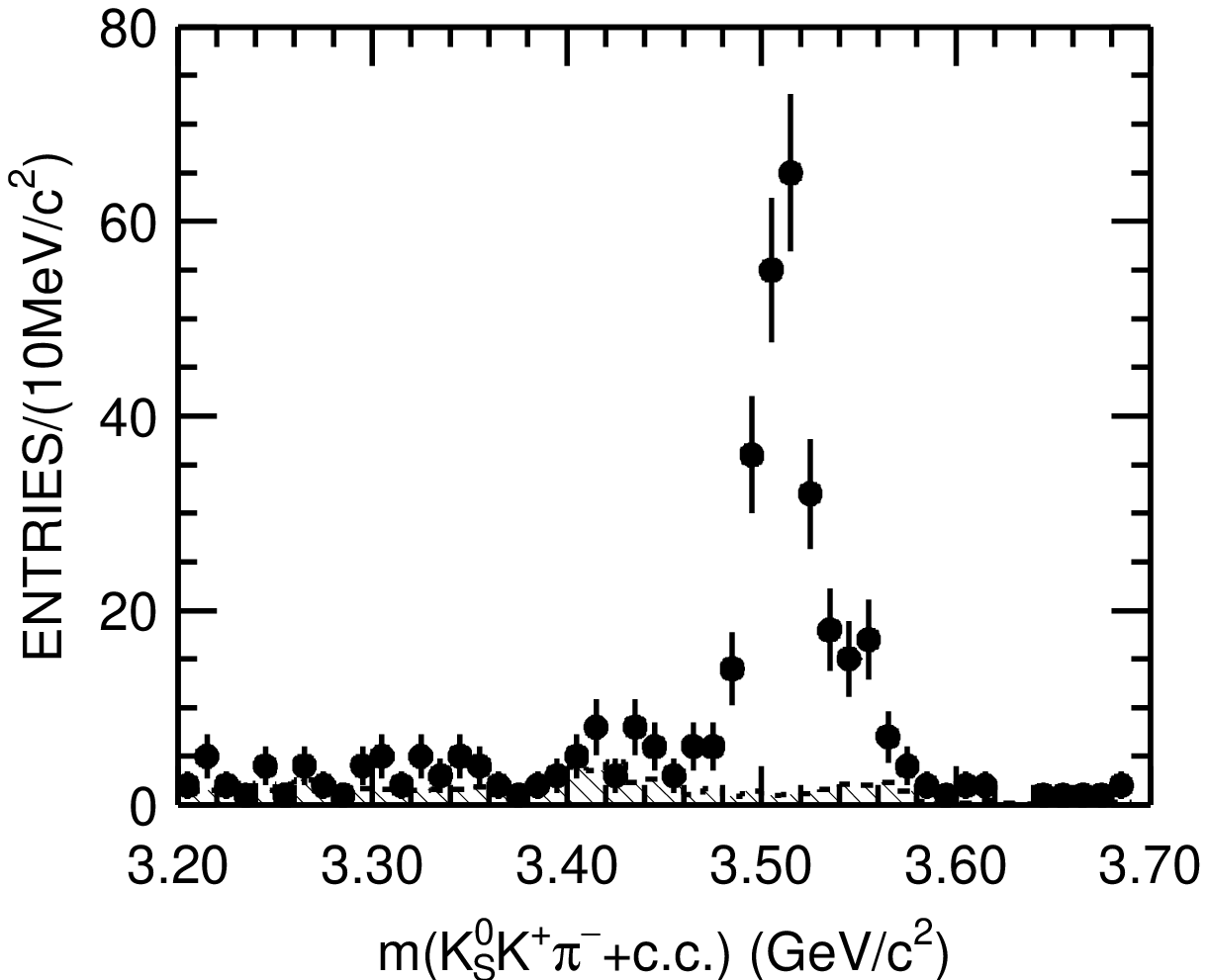,width=0.45\textwidth,height=5.0cm}}}
\caption{Invariant mass distributions for $\psip\ra\kskp$ candidate
  events in the low mass region (upper plot) and high mass region
  (lower). Dots with error bars are data, and the hatched histogram is
  simulated background.}
\label{kskpi}
\end{figure}

With this selection, Fig.~\ref{kskpi} shows the mass distribution
of $\ks K^+\pi^-$ and $\ks K^-\pi^+$ for candidate events. 
There is a clear $\chico$ signal, 
but no clear $\eta(1405)/\eta(1475)$ signal. 
The biggest background contamination comes from
$\psip\ra\piz\kskp$, which is estimated with the data sample, and the other
backgrounds are estimated by Monte Carlo (MC) simulation. 

\begin{figure}[htb]
\centerline{\hbox{\psfig{
file=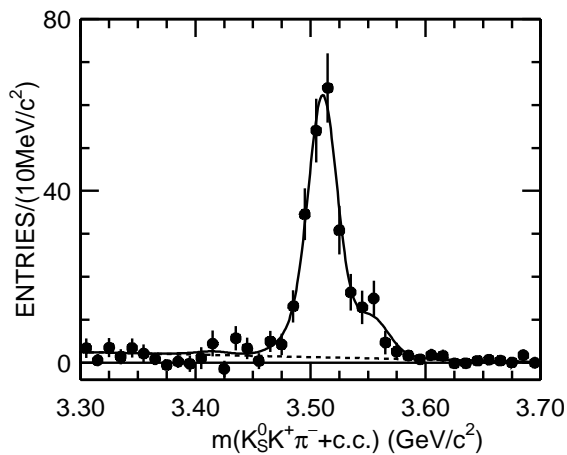,width=0.45\textwidth,height=6cm}}}
\caption{The result of the $\kskp$ mass fit. The curve shows the best fit described in the text.}
\label{kskpimnfit}
\end{figure}

In the high mass region, the fit of the $\kskp$ invariant mass
spectrum is performed after subtracting the known background, and a
second order polynomial is used to describe the shape of the remaining
unknown background (see Fig.~\ref{kskpimnfit}). The $\chicz$ peak is
described with a Breit-Wigner folded with a double-Gaussian resolution
function determined from MC simulation, while the $\chico$ and
$\chict$ peaks are described only with double-Gaussians resolution
functions because their widths are much smaller than the mass
resolution.  The masses of the three $\chicJ$ states and the width of
$\chicz$ are fixed to PDG values~\cite{PDG04}.

A binned maximum likelihood method is used to fit all events with $\ks
K^\pm\pi^\mp$ mass between 3.2 and 3.65 $\gev/c^2$.  The numbers of
events are $3.9\pm4.6$, $220\pm16$, and $28.4\pm 7.6$ with statistical
significances of 0.9$\sigma$, 22.0$\sigma$, and
4.8$\sigma$~\cite{sigma} for $\chicz$, $\chico$, and $\chict$,
respectively.


\begin{figure}[h]
\centerline{\hbox{\psfig{file=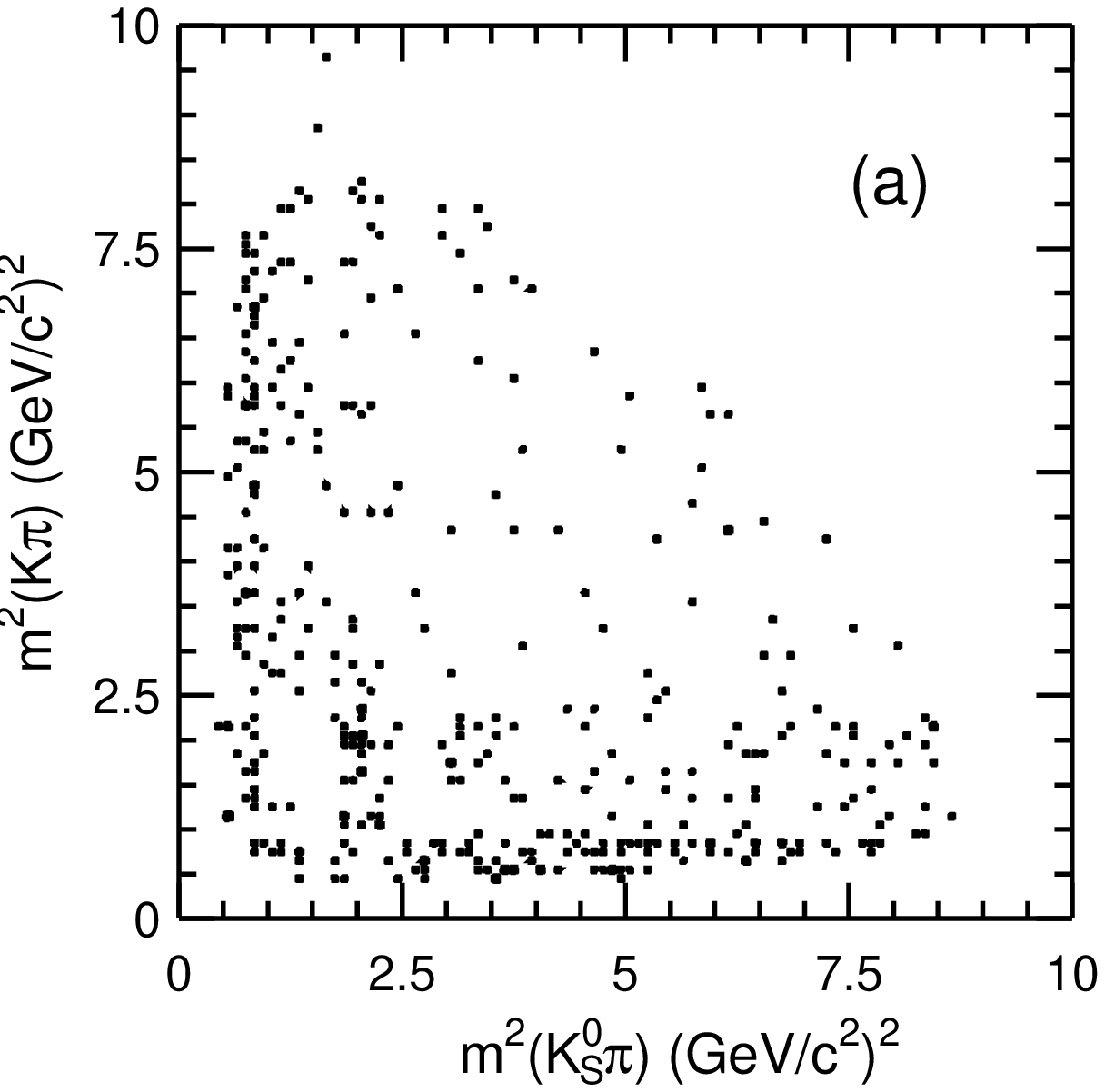,width=0.4\textwidth}}}
\centerline{\hbox{\psfig{file=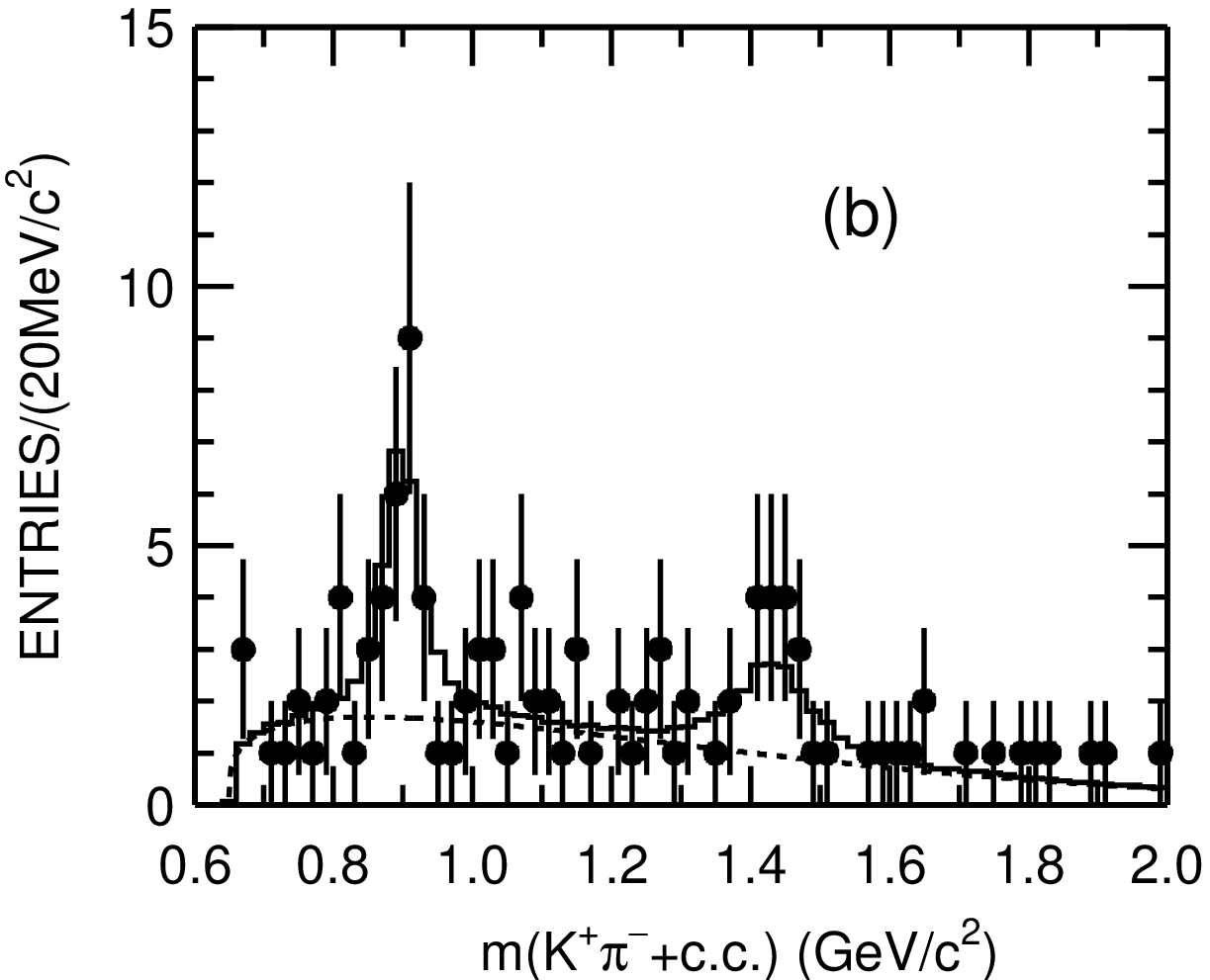, width=0.45\textwidth,height=6cm}}}
\centerline{\hbox{\psfig{file=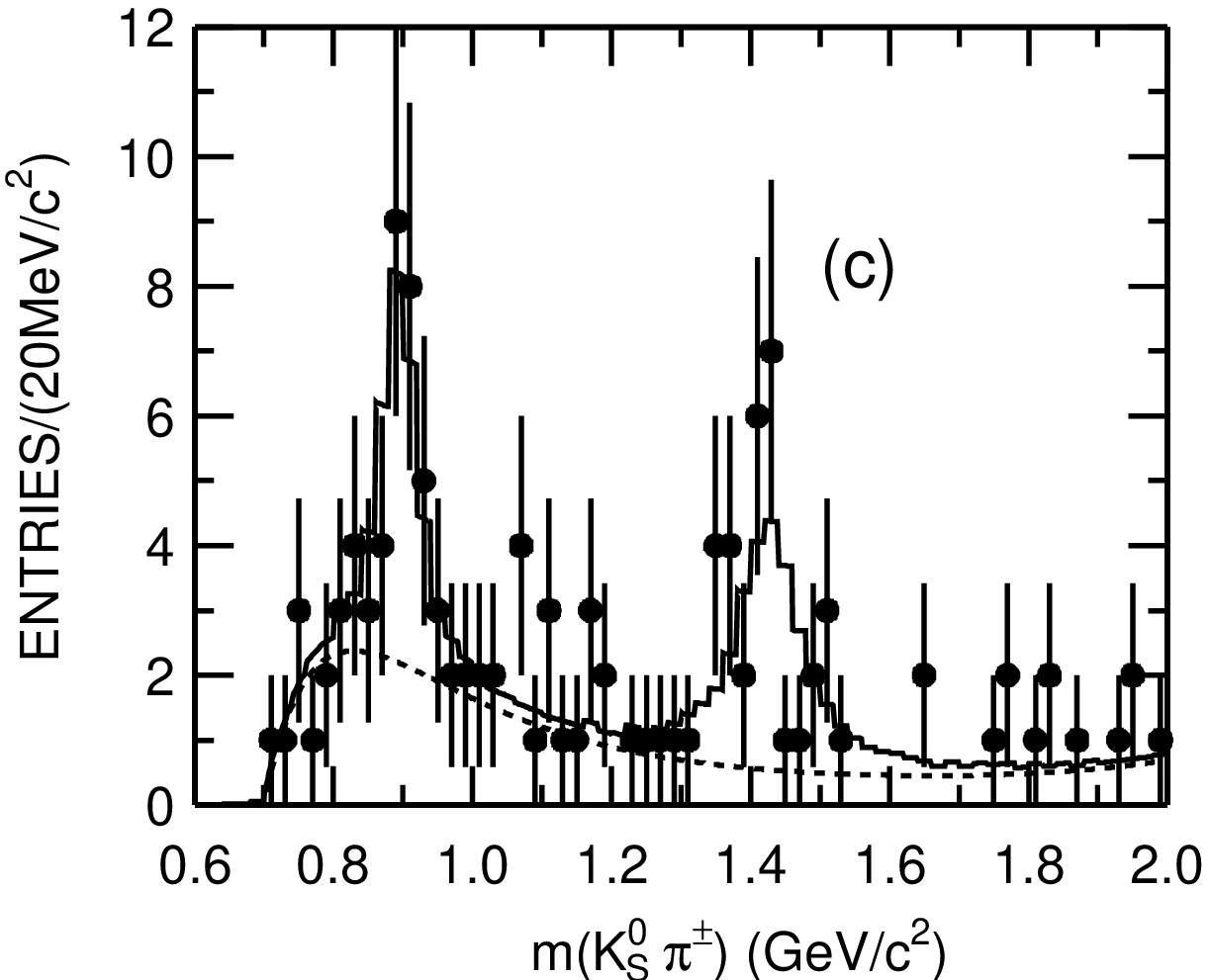, width=0.45\textwidth,height=6cm}}}
\caption{(a) Dalitz plot of $\chico\to\kskp$ candidate events. The (b)
  $K^\pm\pi^\mp$ and (c) $\ks\pi^\pm$ invariant mass distributions. In
  (b) and (c), dots with error bars are data, and the histograms show
  the best fits described in the text.}
\label{chic1dalitz}
\end{figure}

Figure~\ref{chic1dalitz}(a) shows the Dalitz plot of $\chico\to\kskp$
candidate events with $3.48\gev/c^2<m_{\ks K^\pm \pi^\mp}<3.53
\gev/c^2$.  The clusters of events indicate $K^*(892)$ and
$K^*_J(1430)$ signals. Figure~\ref{chic1dalitz}(b) shows the
$K^\pm\pi^\mp$ invariant mass distribution after an additional requirement
$m_{\ks\pi^\pm}>1.0\gev/c^2$ to reject $K^{*\pm}K^\mp$ events.
Figure~\ref{chic1dalitz}(c) shows the $\ks\pi^\pm$ invariant mass
after the requirement $m_{K^\pm\pi^\mp}>1.0\gev/c^2$ to
reject $K^{*0}\overline{K}^0$ events.

For $\chico\to\kskp$ candidate events, the $K^\pm\pi^\mp$ and
$\ks\pi^\pm$ mass spectra are fitted with $K^*(892)$ and $K^*_J(1430)$
signal shapes determined from MC simulations plus a threshold function
for background.  For $K^*(892)^0$ and $K^*(892)^\pm$, the fitted
numbers of events are $22.5\pm7.3$ and $26.7\pm11.0$ with
corresponding statistical significances of $3.5\sigma$ and $3.0\sigma$,
respectively.  For $K^*_J(1430)$, there are three states:
$K^*_2(1430)$, $K^*_0(1430)$, and $K^*(1410)$ around 1430
$\mev/c^2$.
With the detection efficiencies averaged with equal weight, the
numbers of events including these three hypotheses are calculated to
be $22\pm15$ and $45\pm26$ for $K^*_J(1430)^0$ and $K^*_J(1430)^\pm$,
respectively~\cite{evtnum}. The upper limits on the numbers of events
at the 90\% C.L.  are calculated to be 41 and 79.

In the low mass region, no $\eta(1405)/\eta(1475)$ signal is observed
in the $\ks K^\pm\pi^\mp$ invariant mass distribution.  Here the fit
is performed under two hypotheses: one for $\eta(1405)$ with mass 1410
$\mev/c^2$, width 51 $\mev/c^2$, and mass resolution 7.1 $\mev/c^2$;
the other for $\eta(1475)$ with mass 1476 $\mev/c^2$, width 87
$\mev/c^2$, and mass resolution 7.7 $\mev/c^2$.  The $\kskp$ invariant
mass distribution is fitted with a Breit-Wigner folded with a Gaussian
resolution and a second order polynomial for background. The mass,
width, and mass resolution are fixed to the values above. The signal
is very weak, so upper limits on the number of events are calculated
to be 11 and 16 for $\eta(1405)$ and $\eta(1475)$, respectively.
\subsection{\boldmath $\psip\rar \gamma K^+K^-\pi^0$}
\begin{figure}[htbp]
\centerline{\hbox{\psfig{file=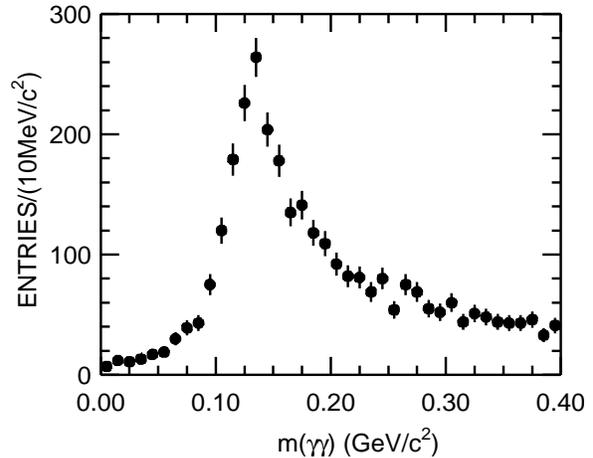,height=6.0cm, width=0.45\textwidth}}}
\caption{The $\gamma\gamma$ invariant mass distribution for $\psip\ra\gamma\gamma\gamma\kk$ candidate events.}
\label{mpiz-dt}
\end{figure}

For this channel, candidate events are required to have two charged
tracks with net charge zero and three photon candidates.  A 4C-fit is
performed under the $\psip\rar\gamma\gamma\gamma K^+K^-$ hypothesis,
and the $\chi^2$ of the fit is required to be less than 15. The
invariant mass of the charged kaon tracks is required to be less than
3.0 GeV/c$^2$ to veto $\psip\rar neutral+\jpsi$ background. With
three selected photons, there are three possible combinations to
reconstruct $\piz$, and the combination with invariant mass closest to
$m_{\piz}$ is taken as the $\piz$ candidate.  Figure~\ref{mpiz-dt}
shows the $\gamma\gamma$ invariant mass distribution, where a clear
$\piz$ signal is observed.

\begin{figure}[htbp]

\centerline{\hbox{\psfig{file=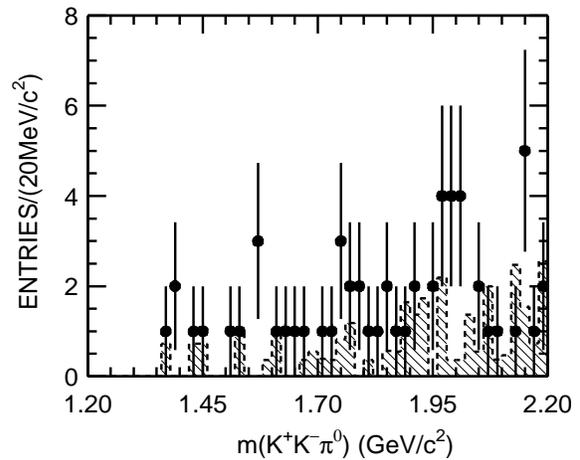,width=0.45\textwidth,height=6cm}}}
\caption{The $\kk\piz$ invariant mass distribution for $\psip\ra\gamma\kk\piz$
 candidate events. Dots with error bars are data, and the hatched histogram 
is the simulated background.}
\label{mkkpiz-dt}
\end{figure}

After requiring $|m_{\gamma\gamma}-m_{\piz}|<0.03\gev/c^2$, 
Fig.~\ref{mkkpiz-dt} shows the  $\kk\piz$ mass distribution
for candidate events. There is no $\eta(1405)/\eta(1475)$ signal in the
low mass region. 
Upper limits on the number of events are calculated to be 9 and 9
for $\eta(1405)$ and $\eta(1475)$, respectively.


\subsection{\boldmath $\psip\rar\gamma\eta\pipi$}\label{getapipi}
The final state of this channel is $\pipi\gamma\gamg$. Events with two
charged tracks with net charge zero and three photon candidates are
selected. A 4C-fit is performed for the hypothesis $\psip\rar
\pipi\gamma\gamg$, and the $\chi^2$ of the fit is required to be less
than 15. Background from $\psip\rar\pipi J/\psi$ is rejected with the
requirement $|m^{\pi^{+}\pi^{-}}_{recoil}-3.1|>0.05$ GeV/$c^2$.
Background from $\psip\rar neutrals + J/\psi$ is suppressed with the
requirement $m_{\gamma\pipi}<2.8\GeV/c^2$, where $m_{\gamma\pipi}$ is
the invariant mass of the $\pipi$ and the photon which does not come
from $\eta$ decay.

\begin{figure}
\includegraphics[height=6.0cm,width=0.45\textwidth]{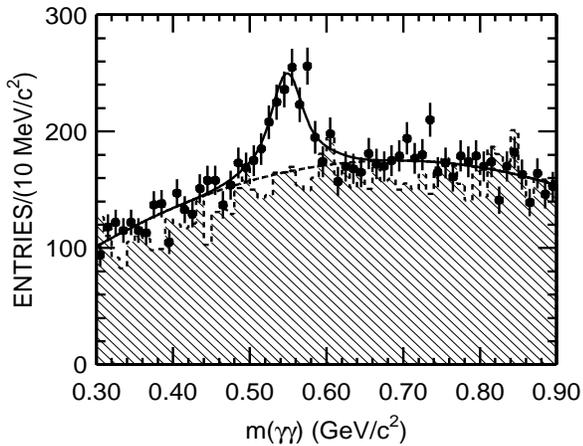}
\caption{\label{mgg} The $\gamg$ invariant mass distribution for
 $\psipto\pi^+\pi^-\gamg\gamma$ candidate events
(dots with error bars). 
The curves show the best fit described in the text. 
The hatched histogram is the $\gamg$ distribution of background events 
from the continuum and the 14M inclusive decay MC sample 
with signal events removed.}
\end{figure}

With the above selection, Fig.~\ref{mgg} shows the $\gamg$ invariant
mass distribution, where $\gamg$ includes all possible combinations
among the three photon candidates. A clear $\eta$ signal is observed.
The smooth background comes from many channels and can be described by
the sum of continuum events and $\psip$ inclusive decay MC events,
where the signal events have been removed and some known background
channels are replaced by MC simulated results. The main background of
the $\eta$ signal comes from $\psipto\eta\pipi\pi^0$, which is
estimated using $\psip$ data.  We also studied other possible channels
listed in the PDG~\cite{PDG04} that might contaminate the $\eta$
signal, but the contamination is negligible.  A fit of the $m_{\gamg}$
spectrum yields 553$\pm$60 events, and the background contamination is
estimated to be 135$\pm$59 by fitting the hatched histogram in
Fig.~\ref{mgg}. In the fit, the $\eta$ signal is described by the
double-Gaussian shape determined from $\psipto\gamma\eta\pipi$ MC
simulation.  After the background contamination is subtracted, the
number of $\eta$ events is 418$\pm$60, with a statistical
significance of 7.3$\sigma$.  Here the background contamination is
subtracted from the total number of observed events,
and the uncertainty on the number of background events is taken as a
systematic error. This method to deal with the background
contamination is also applied to the following analyses.

An $\eta$ candidate is defined with the requirement
$|m_{\gamg}-0.548|<0.04\GeV/c^2$.  Figure~\ref{fit2} shows the
$\eta\pipi$ invariant mass distributions in the low and high mass regions.
Clear $\etap$ and $\chi_{c1}$ signals are seen.

\subsubsection{$\psip\rar\gamma\eta^\prime(958)$ and 
$\gamma\eta(1405)/\eta(1475)$}


\begin{figure}[hbt]
\includegraphics[height=6.0cm, width=0.45\textwidth]{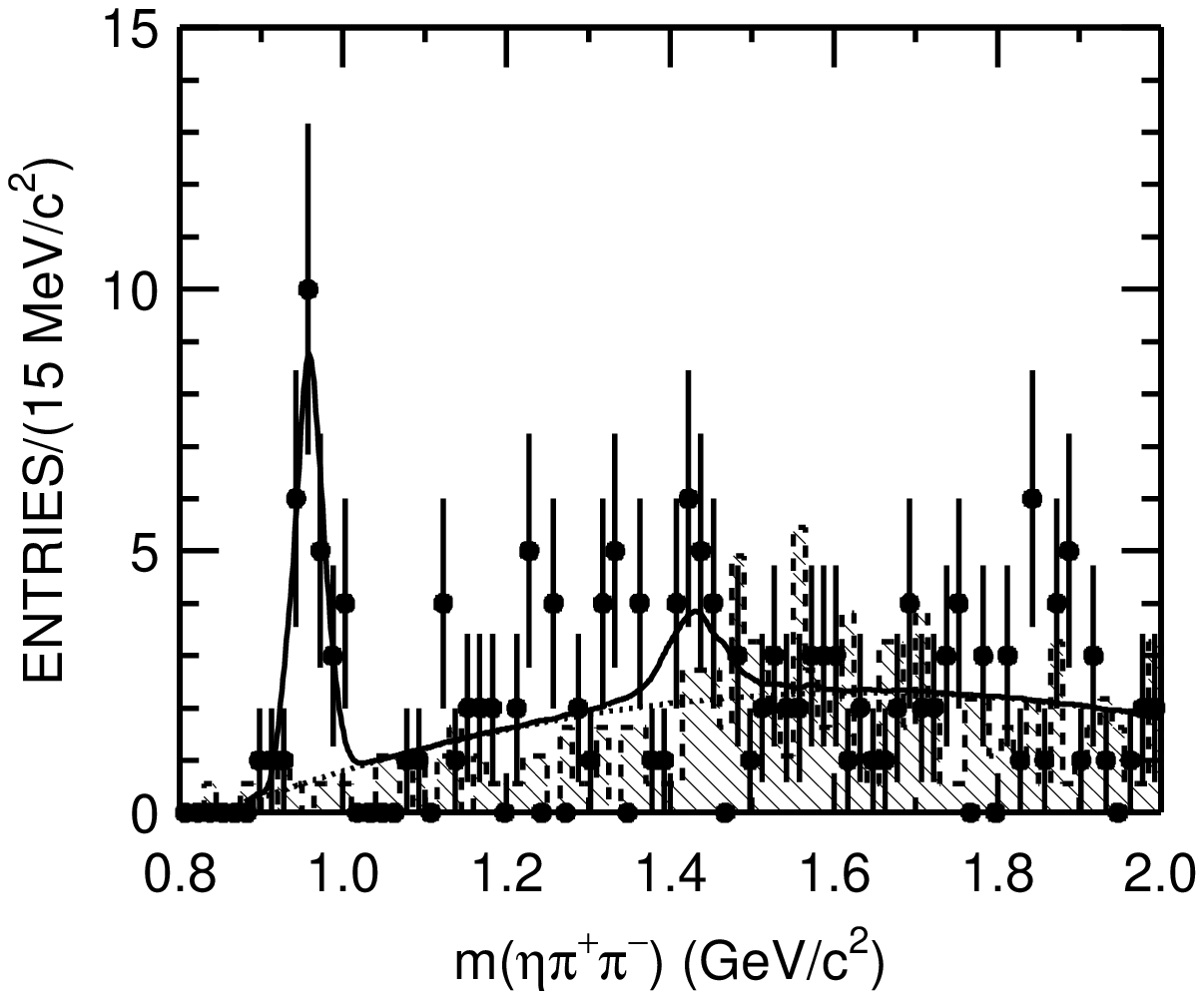}
\includegraphics[height=6.0cm, width=0.45\textwidth]{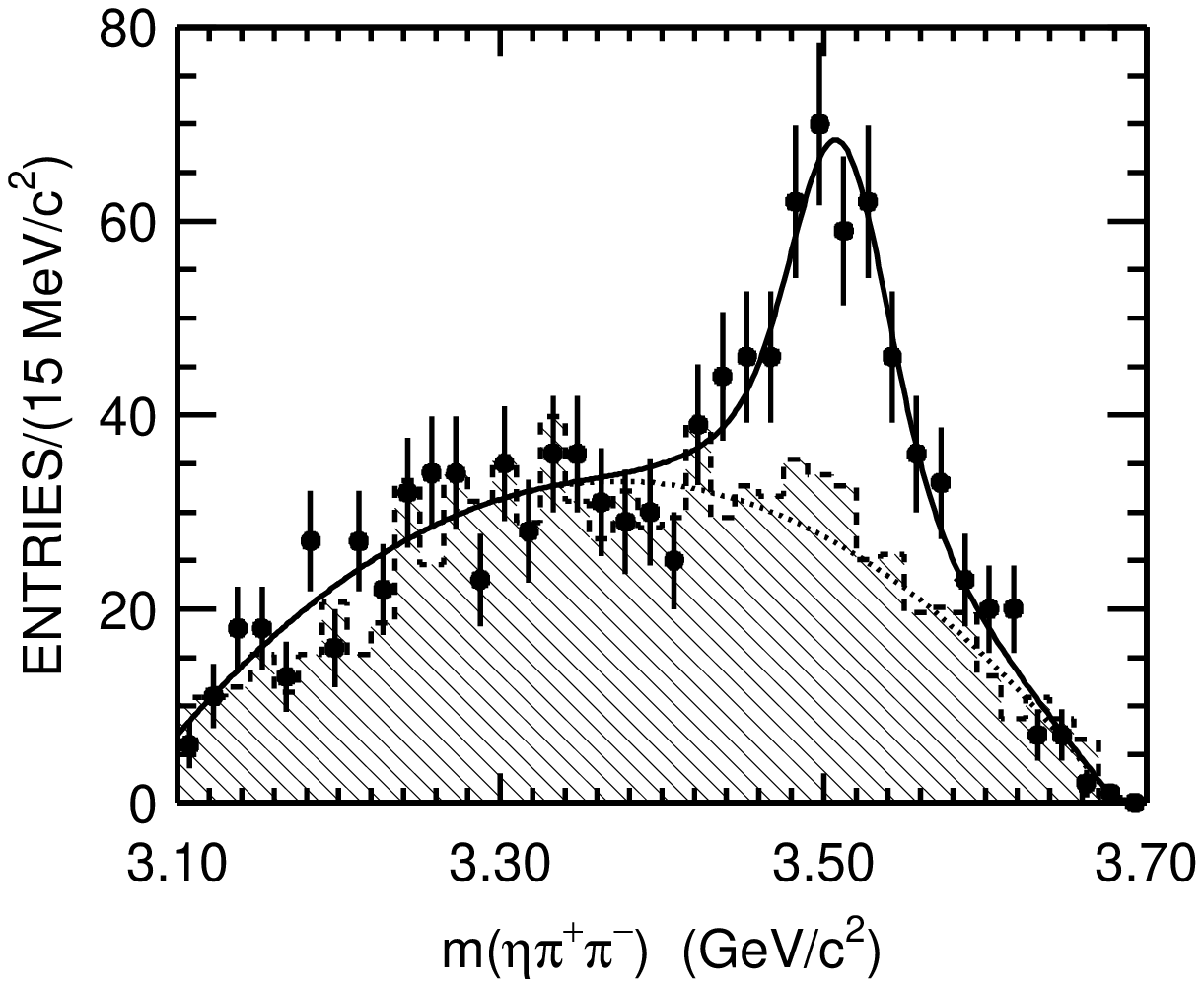}
\caption{\label{fit2} The $\eta\pipi$ invariant mass distributions for
$\psipto\gamma\eta\pipi$ candidate events in the low mass
 region (upper plot) and high mass region (lower). The dots with error bars
are data, and the hatched histogram is the background estimated from
the $\eta$ sidebands. The curves show the best fit described in the text.}
\end{figure}

Besides the $\eta^\prime(958)$ signal, there is also a small peak at
1430$\mev/c^2$, which could be an $\eta(1405)$, $f_1(1420)$,
$\rho(1450)$, or $\eta(1475)$ listed by PDG~\cite{PDG04}. $f_1(1420)$
and $\eta(1475)$ dominantly decay into $K\overline{K}\pi$, but no
significant signal of $f_1(1420)$ or $\eta(1475)$ is observed in the
$K_{S}K\pi$ invariant mass distribution (see Fig. \ref{kskpi}(a)). The
 $\psipto\gamma\rho(1450)$ decay is forbidden by C-parity
conservation.  So the peak at 1430$\mev/c^2$ is assumed to be
$\eta(1405)$ signal, and more will be discussed later.

Assuming $\etap$ and $\eta(1405)$ signals, 
the low mass region is fitted with the MC distributions 
plus a second order polynomial for background (see Fig. \ref{fit2}). 
The fit yields 24.2$\pm$5.4 and 13.8$\pm$7.0 events, and the peaking background
events are estimated to be 0.9$\pm$1.4 and 4.0$\pm$4.5 
from $\eta$ sidebands. The $\eta$ sideband region is
defined by  $|m_{\gamg}-0.38|<0.04\GeV/c^2$ and $|m_{\gamg}-0.72|<0.04\GeV$.
After background subtraction, the numbers of
$\etap$ and $\etafl$ events  become 23$\pm$5 and 10$\pm$7, 
and the  statistical significances are 6.6$\sigma$ and 1.4$\sigma$, 
respectively. 

Since the significance of $\eta(1405)$
is low and there is no clear $\eta(1475)$ signal, upper limits 
at the 90\% C.L. on the numbers of events for  
$\eta(1405)$ and $\eta(1475)$ are calculated 
 to be 24 and 20, respectively.

\subsubsection{$\psip\rar\gamma\chi_{c1}$}

The fit in the high $m_{\eta\pipi}$ region yields 
256$\pm$28 $\chi_{c1}$ events (see Fig. \ref{fit2}), and 
the peaking background events are estimated to be 34$\pm$15 
from the $\eta$ sideband region. The $\eta$ sideband region is
defined by  $|m_{\gamg}-0.38|<0.04\GeV/c^2$ and $|m_{\gamg}-0.72|<0.04\GeV$. 
 After the background contamination 
is subtracted, the number of $\chi_{c1}$ signal events 
becomes 222$\pm$28, with 8.8$\sigma$ statistical significance.

\begin{figure}
\begin{center}
\includegraphics[width=0.45\textwidth]{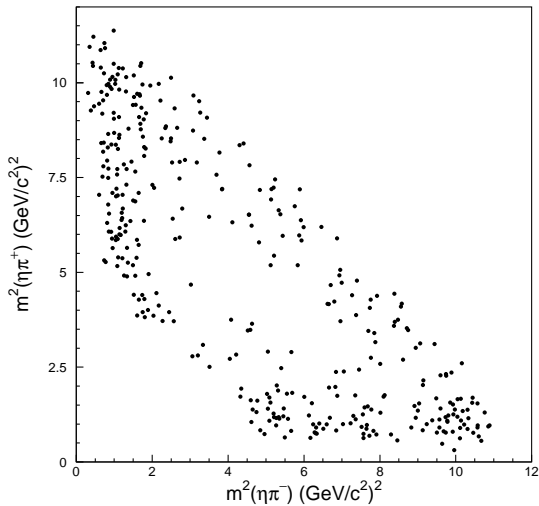}
\caption{\label{scatter1} Dalitz plot of $\chi_{c1}\rar\eta\pipi$
candidate events.}
\end{center}
\end{figure}

The Dalitz
plot of $\chi_{c1}\rar\eta\pipi$ candidate events within the
$\chi_{c1}$ mass window (3.46-3.56)$\gev/c^2$ is shown in
Fig.~\ref{scatter1}.  The horizontal and vertical clusters with
$m_{\eta\pi}$ around 1$\gev/c^2$ correspond to $\chi_{c1}\rar
\aon\pi$, and the diagonal band is $\chi_{c1} \rar f_2(1270)\pi$.

\begin{figure}[h!]
\includegraphics[height=6.0cm,width=0.45\textwidth]{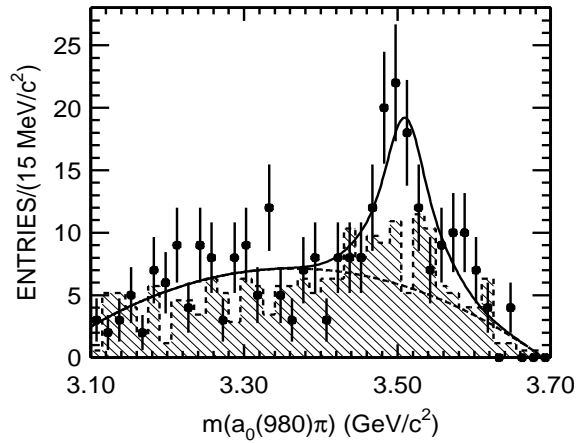}
\caption{\label{ma0pi-fit}The $\api$ invariant mass distribution 
  for $\psipto\gamma \aon^{\pm}\pi^{\mp}$ candidate events (dots with
  error bars). The curves show the best fit described in the text.
  The hatched histogram is the $m_{\eta\pipi}$ distribution of the
  events in $\aon$ sideband region.}
\end{figure}

The $\aon^{\pm}\pi^{\mp}$ invariant mass distribution for events
satisfying $(|m_{\eta\pi^{\pm}}-0.985|<0.1\GeV/c^2)$ is shown in Fig
\ref{ma0pi-fit}. The distribution is fitted with a MC determined
double-Gaussian function plus a second order polynomial for the background.  
The fit yields 79$\pm$14 $\chi_{c1}$ candidate events, and the number of
background events contributing to the peak is estimated to be
21$\pm$11 by using a similar fit for events from the $\aon$ sideband
region.  The $\aon$ sideband region is
defined by  $|m_{\eta\pi^{\pm}}-1.6|<0.3\GeV/c^2$. 
After subtraction, the number of $\chi_{c1}$ signal events is
determined to be 58$\pm$14, with a 4.5$\sigma$ statistical
significance.

\begin{figure}
\begin{center}
\includegraphics[width=0.45\textwidth]{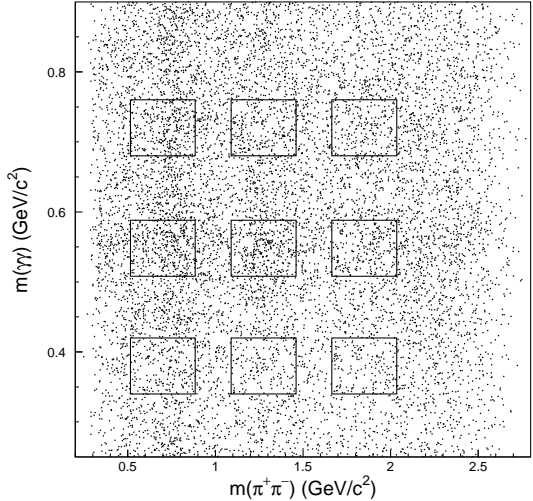}
\caption{\label{mpipi-meta-dt} The scatter plot of $m_{\gamg}$ versus
  $m_{\pipi}$ for candidate events. }
\end{center}
\end{figure}
\begin{figure}[h!]
\includegraphics[height=6.0cm,width=0.45\textwidth]{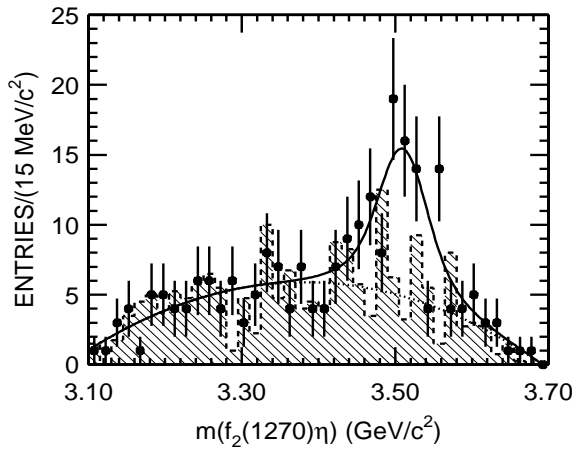}
\caption{\label{mf2eta-fit} Invariant mass distribution
  of $\ft\eta$ for $\psipto\gamma\ft\eta$ candidate events (dots with
  error bars). The curves show the best fit described in the text.
  The hatched histogram is the $m_{\gamg\pipi}$ distribution of events
  in the sideband regions.}
\end{figure}

Figure~\ref{mpipi-meta-dt} shows the scatter plot of $m_{\gamg}$
versus $m_{\pipi}$ of the $\chi_{c1}$ candidate events. The $\ft\eta$
signal region is defined by $|m_{\pipi}-1.275|<0.185\GeV/c^2$ and
$|m_{\gamg}-0.548|<0.04\GeV/c^2$.  The $\ft$ sideband region is
defined by $|m_{\pipi}-0.7|<0.185\GeV/c^2$ and
$|m_{\pipi}-1.85|<0.185\GeV/c^2$, and the $\eta$ sideband region by
$|m_{\gamg}-0.38|<0.04\GeV/c^2$ and $|m_{\gamg}-0.72|<0.04\GeV$ (see
Fig. \ref{mpipi-meta-dt}). The central box in the figure is the signal
region, the two boxes located above and below the signal region are the $\eta$ sidebands (named
$B_{\eta}^{1}$ and $B_{\eta}^{2}$), and the two on the left and right
of the signal region are the $\ft$ sidebands (named
$B_{f_2}^{1}$ and $B_{f_2}^{2}$).  The four boxes at the corners are
used to estimate the phase space contribution (named $B_{ph}^{i},
i=1,2,3,4$).  We use the formula:
$$\frac{1}{2}\times (B_{\eta}^{1} + B_{\eta}^{2} +  
B_{f_2}^{1} +  B_{f_2}^{2}) - \frac{1}{4}\times
(B_{ph}^{1} + B_{ph}^{2} +B_{ph}^{3} + B_{ph}^{4})$$
to obtain the $m_{\eta\pipi}$ distribution of the events in 
the sideband regions.
Figure \ref{mf2eta-fit} shows the  $\eta\pipi$ invariant mass distributions of
the signal and background regions, and 
a clear $\chi_{c1}$ signal is observed. 

The $\chi_{c1}$ signal is fitted with a double-Gaussian function
determined from MC simulation plus a second order polynomial to describe the
background (see Fig. \ref{mf2eta-fit}). The fit yields 65$\pm$13
events, and fitting the sideband region events yields 12$\pm$7
sideband background events. After subtraction, the number of
$\chi_{c1}$ signal events is 53$\pm$13 with a 4.8$\sigma$ statistical
significance.

$\xc$ decays to $\api$ and $\ft\eta$ yield
the same final state $\eta\pipi$.
MC studies show that the sideband analysis described above separates
the two channels without cross contamination.
However, any interference effects are not taken into consideration 
because of the low statistics.


\section{\boldmath Simulation and efficiency }
Monte Carlo simulation is used for mass resolution and detection 
efficiency determination.
In this analysis, a GEANT3 based Monte Carlo package with detailed
consideration of the detector performance (such as dead electronic
channels) is used.  The consistency between data and Monte Carlo has
been carefully checked in many high purity physics channels, and the
agreement is reasonable~\cite{SIMBES}.
 
For $\pspto \gamma\etap$ and $\pspto \gamma\eta(1405/1475)$, the 
photons are distributed according to a $1+\cos^2\theta$ distribution. 
For $\pspto \gamma\chicj$ under 
the assumption that the processes are pure E1 transitions~\cite{E1transition},
 the photons  are generated as $1+\cos^2\theta$, 
~$1-\frac{1}{3}\cos^2\theta$, and ~$1+\frac{1}{13}\cos^2\theta$ for $\chicz$, 
$\chico$, and $\chict$, respectively. Multihadronic decays of $\etap$, 
$\eta(1405/1475)$, and $\chicj$ are simulated using phase space distributions.

In the MC simulation for $\psipto\gamma\chi_{c1}$, $\chi_{c1}\rar\api$,
 the width of $a_0(980)$ is assumed to be 75$\mev/c^2$ in the
 determination of the 
detection efficiency. 
The uncertainty of the efficiency due to the uncertainty of
the $a_0(980)$ width is taken 
as a systematic error in the branching 
fraction of $\chi_{c1}\rar\api$.

The efficiencies for the determination of the branching fractions of
$\psipto\gamma\eta\pipi$, $\chi_{c1}\rar\eta\pipi$ and
$\chi_{c1}\rar\kskp$ are determined from a weighted average over the
intermediate processes.

\section{\boldmath Systematic errors}

\begin{table*}[htbp]
\caption{Summary of systematic errors (\%). MDC, 4C, PID, $\gamma$eff.,
$N_{\psip}$, Int., MC, Bg., Fit. and $\ks$ rec. are for tracking, kinematic fit, particle
identification, $\gamma$ detection efficiency, $\psip$ total number,
the branching fractions of the intermediate states, MC statistics, background, fitting on mass 
spectrum and $\ks$ reconstruction, respectively.}
\begin{tabular}{l|ccccccccccc}\hline\hline
 Channel ($\psipto$)      & MDC &4C& PID & $\gamma$ eff.& $N_{\psp}$ & Int. &MC  
& Bg.& Fit. & $\ks$ rec.& Total\\\hline
$\gamma\chicz\to\gamma\kskp$ &8.0&6.0&-&2.0&4.0&8.1&1.3&20.6&18.0&3.4&30.8\\
$\gamma\chico\to\gamma\kskp$ &8.0&6.0&-&2.0&4.0&9.5&1.3&0.2&0.7&3.4&15.0\\
$\gamma\chict\to\gamma\kskp$ &8.0&6.0&-&2.0&4.0&9.4&1.3&2.9&3.9&3.4&15.7\\\hline

$\gamma\chico\to \gamma K^*(892)^0\ks$ &8.0&6.0&-&2.0&4.0&9.5&1.3&-&1.9&3.4&15.1\\
$\gamma\chico\to \gamma K^*(892)^\pm K^\mp$ &8.0&6.0&-&2.0&4.0&9.5&1.3&-&3.2&3.4&15.3\\
$\gamma\chico\to \gamma K^*_J(1430)^0\ks\to \gamma\kskp$ &8.0&6.0&-&2.0&4.0&9.5&1.3&-&5.9&3.4&16.1\\
$\gamma\chico\to \gamma K^*_J(1430)^\pm K^\mp\to\gamma \kskp$ &8.0&6.0&-&2.0&4.0&9.5&1.3&-&15.2&3.4&21.4\\ \hline
$\gamma\eta(1405)\ra\gamma\kskp$&8.0&6.0& -  &2.0&4.0&-&1.5&-&5.0&3.4&12.6\\
$\gamma\eta(1475)\ra\gamma\kskp$&8.0&6.0& -  &2.0&4.0&-&1.5&-&24.0&3.4&26.7\\
$\gamma\eta(1405)\ra\gamma\kk\piz$ &4.0&4.0&2.0&6.0 &4.0&-&1.7&-&12.3&-&15.6\\
$\gamma\eta(1475)\ra\gamma\kk\piz$ &4.0&4.0&2.0&6.0 &4.0&-&1.7&-&20.0&-&22.2 
\\ \hline
$\gamma\eta\pipi$& 4.0 & 4.0 & 2.0 & 6.0  & 4.0 & 0.7 & 2.3 & 14.0
   & 7.7    &-& 18.7 \\     
$\gamma\etap$    & 4.0 & 4.0 & 2.0 & 6.0  & 4.0 & 3.5 & 1.6 & 6.0 
   & 2.9  &-& 12.1 \\      
$\gamma\etafl\rar\gamma\eta\pipi$ & 4.0 & 4.0 & 2.0 & 6.0  & 4.0 & 0.7 
& 1.8 & -  & 10.9 &-& 14.5 \\
$\gamma\eta(1475)\rar\gamma\eta\pipi$& 4.0 & 4.0 & 2.0 & 6.0  & 4.0 & 0.7 
& 2.0 & - & 10.0 &-& 13.9 \\ \hline
$\gamma\chico\to\gamma\eta\pipi$ & 4.0 & 4.0 & 2.0 & 6.0 & 4.0 & 9.5 
& 1.1  & 6.7  & 6.4  &-&  16.3 \\ 
$\gamma\chico\to\gamma \aon^+\pi^- +c.c.$ & 4.0 & 4.0 & 2.0 & 6.0  & 4.0 
& 14.4 & 1.0 & 19.2  & 6.2  &-& 26.5 \\ 
$\gamma\chico\to\gamma\ft\eta$ &  4.0 & 4.0 & 2.0 & 6.0  & 4.0 & 10.0
 & 1.0 & 13.2    & 2.8  &-& 19.3 \\ \hline \hline
\end{tabular}
\label{Tab:sys}
\end{table*}

Many sources of systematic error are considered. 
Systematic errors associated with
the MDC tracking, 
kinematic fitting, particle identification, 
 and photon selection efficiencies are determined by comparing $J/\psi$
  and $\psi(2S)$ 
data and Monte Carlo simulation for pure data samples, 
such as $\psi(2S) \rar \pi^+ \pi^- J/\psi$.

The uncertainties on the total number of $\psi(2S)$ 
events, 
the branching fractions of intermediate states, the $\aon$ 
width, the detection efficiency, the background 
contamination, and the fitting on the mass spectrum
are also considered as
systematic errors.  Table~\ref{Tab:sys} summarizes
the systematic errors for all channels.
\section{\boldmath Results and Discussion}

\begin{table*}
\begin{center}
\caption{ \label{Tab-psip} Measured branching fractions and upper limits 
(90\% C.L.) for $\psi(2S)$ decays. Results for corresponding
$J/\psi$ decays~\cite{PDG04} and the
ratio $Q_h=\frac{\BR(\psi(2S)\rar h)}{\BR(J/\psi\rar h)}$ are also given.}
\begin{tabular}{cccccc} \hline \hline
Channel ($\psipto$)  & $n^{sig.}$ & $\eff$  (\%)&
$\BR_{\psipto}(\times 10^{-4})$ & $\BR_{\jpsito}(\times 10^{-4})$ & $\frac{\BR(\psip)}{\BR(\jpsi)}$(\%) \\ \hline
${\gamma\eta\pipi}^{a}$ & 418$\pm$60  & 8.69 
& 8.71$\pm$1.25$\pm$1.64 &  $-$ & $-$ \\
${\gamma\eta\pipi}^{b}$ & $-$  & $-$  
& 3.60$\pm$1.42$\pm$1.83 &  39$\pm$7.3~\cite{markiii} & 9.2$\pm$6.2 \\
$\gamma\etap$      & 23$\pm$5 & 7.58 
& 1.24$\pm$0.27$\pm$0.15  & 43.1$\pm$3  & 2.9$\pm$0.7 \\
$\gamma\eta(1405)\rar\gamma\eta\pipi$ & 10$\pm$7 & 5.06 
& 0.36$\pm$0.25$\pm$0.05   & 3.0$\pm$0.5 & 12$\pm$10  \\
& $<24$ & 5.06 & $<1.0$ & 3.0$\pm$0.5 & $<33$ \\
$\gamma\eta(1475)\rar\gamma\eta\pipi$ & $<20$ & 4.80 
& $<0.83$ & 3.0$\pm$0.5 & $<28$ \\ \hline
$\gamma\eta(1405)\to \gkkp\ ^c$ & $<11$ &4.54
& $<0.8$ & $28\pm6$ & $<2.9$\\
$\gamma\eta(1475)\to \gkkp\ ^c$ & $<16$ &4.58 
& $<1.5$ & $28\pm6$ & $<5.4$ \\\hline 
$\gamma\eta(1405)\to \gkkp\ ^d$ & $<9$ &3.63 
& $<1.3$ & $28\pm6$ & $<4.6$ \\
$\gamma\eta(1475)\to \gkkp\ ^d$  & $<9$ &3.54 
& $<1.4$ & $28\pm6$ & $<5.0$ \\ 
\hline \hline
\multicolumn{6}{l}{$^a$ all processes in the $\psipto\gamma\eta\pipi$;} \\
\multicolumn{6}{l}{$^b$ all processes excluding $\psipto\gamma\chi_{c1}\rar\gamma\eta\pipi$;}\\
\multicolumn{6}{l}{$^c$ the decay mode is $\gamma\kskp$;}\\ 
 \multicolumn{6}{l}{$^d$ the decay mode is $\gamma\kk\piz$.}
\end{tabular}
\end{center}
\end{table*}

Tables~\ref{Tab-psip},~\ref{sumxcj}, and \ref{Sumxc1} summarize the
results for the channels measured in this analysis.  Table
\ref{Tab-psip} lists the branching fractions of $\psip$ decays. To
compare with the 12\% rule, Table \ref{Tab-psip} also includes the
corresponding $\jpsi$ branching fractions~\cite{PDG04}, as well as the
ratio $Q_h$ of $\psip$ to $\jpsi$ branching fractions for each
channel.  Decay of $\psip$ to $\gamma\eta\pipi$ is consistent with the
12\% rule expectation within errors; decays of $\psip$ to
$\gamma\eta(1405)\rar\eta\pipi$ and $\gamma\eta(1475)\rar\eta\pipi$
cannot be tested because of low statistics; while the other modes are
suppressed by a factor of $2\sim4$.
The $\psipto\gamma\etap$ branching fraction with $\etap\rar\eta\pipi$
is  more precise than $(2.00\pm0.59\pm0.29)\times10^{-4}$ measured
by BESI~\cite{getap}.

No signal for $\eta(1405)/\eta(1475)$ is observed in either $\gamma\kskp$ or 
$\gamma\kk\piz$ final states. There is a small peak at 1430$\mev/c^2$ 
in the $\gamma\eta\pipi$ final state, and we have treated it as $\eta(1405)$ signal. Because of its low statistics, we also
set the upper limit at the 90\% C.L. for 
$\psipto\gamma\eta(1405)/\eta(1475)\to \gamma\eta\pipi$.
As shown in Table \ref{Tab-psip},
the upper limits at the 90\% C.L. on
$\psipto\gamma\eta(1405)/\eta(1475)\to \gkkp$ and $\gamma\eta\pipi$
are at the same level $0.8\sim2.0\times10^{-4}$.

\begin{table*}
\begin{center}
\caption{\label{sumxcj}  Branching fractions for $\chicj\to\kskp$ and
$\chicj\to\eta\pipi$. Here 
$\BR(\psip\ra\gamma\chicz)=(8.6\pm0.7)\%$, 
$\BR(\psip\ra\gamma\chico)=(8.4\pm0.8)\%$ and 
$\BR(\psip\ra\gamma\chict)=(6.4\pm0.6)\%$ are used in the calculation.}
\begin{tabular}{c|c|c|c|c|c}\hline\hline
Channel & $\chi_{cJ}$ & $n^{sig.}$&$\eff$ (\%)& $\BR(\times 10^{-3})$&BESI $(\times10^{-3})$\\\hline
         & $\chicz$ &$<10$& 6.24 & $<0.3$& $<0.71$\\
$\kskp$  & $\chico$ &$220\pm16$& 6.80&$4.1\pm0.3\pm0.7$ 
                    & $2.46\pm0.44\pm0.65$\\
         & $\chict$ &$28.4\pm7.6$& 5.82 & $0.8\pm0.3\pm0.2$& $<1.06$ \\\hline
         & $\chicz$ &$<32$& 6.64 & $<1.2$& -\\ 
$\eta\pipi$ & $\chico$ & 222$\pm$28 & 7.90 & 6.1$\pm$0.8$\pm$1.0 & -\\
         & $\chict$ &$<48$& 7.17 & $<2.2$& - \\ \hline \hline
\end{tabular}
\end{center}
\end{table*}

\begin{table*}
\begin{center}
\caption{\label{Sumxc1} Branching fractions of $\chico\to K^*\kbar$, 
$\aon\pi$ and $\ft\eta$. Here 
$\BR(\psip\ra\gamma\chico)=(8.4\pm0.8)\%$ is used in the calculation.}
\begin{tabular}{c|c|c|c}\hline\hline
Channel & $n^{sig.}$ & $\eff$  (\%)   &  $\BR(\times 10^{-3})$ \\ \hline
$K^*(892)^0\kzbar + c.c.$ & $22.5\pm7.3$  & 7.67 & $1.1\pm0.4\pm0.2$ \\
$K^*(892)^+K^- + c.c.$    & $26.7\pm11.0$ & 6.20 & $1.6\pm0.7\pm0.3$ \\ \hline
$K_J^*(1430)^0\kzbar + c.c.\to\kskp$ & $<41$ & 6.28  & $<1.0$ \\
$K_J^*(1430)^+K^- + c.c.\to\kskp$ & $<79$ & 5.00   & $<2.5$  \\ \hline
$\aon^+\pi^- + c.c.\rar\eta\pipi$ &58$\pm$14 & 6.10 & 2.0$\pm$0.5$\pm$0.5\\
$\ft\eta$ & 53$\pm$13 & 6.55 &  2.1$\pm$0.5$\pm$0.4  \\ \hline \hline
\end{tabular}
\end{center}
\end{table*}

In the above study in the high $\eta\pipi$ mass region, only
$\chi_{c1}$ is considered.  If we fit $m_{\eta\pipi}$ with $\chi_{c0,
  1, 2}$ together, the fit yields $-32\pm28$, 250$\pm$32, and
17$\pm$26 for $\chi_{c0}$, $\chi_{c1}$, and $\chi_{c2}$, respectively.
The difference in the number of $\chi_{c1}$ events is 2.3\%, which has
been taken into account as a systematic error.  The 90\% C.L. upper
limits on the number of $\chi_{c0}$ and $\chi_{c2}$ events are
calculated to be 32 and 48, and the relative systematic errors are 
 12.4\% and 13.3\%, respectively.  The corresponding upper
limits at the 90\% C.L. on the branching fractions are listed in Table
\ref{sumxcj}.

For the $\chicJ\to\kskp$ decays (listed in Table \ref{sumxcj}), 
we get higher precision
results compared to the BESI experiment~\cite{besichic}. The 
branching fraction of $\chico\to\kskp$ is consistent with the 
BESI result within 1$\sigma$, while the results for $\chicJ\to\eta\pipi$ 
decays are all first measurements. The branching fractions of 
$\chico$ decays into intermediate processes listed 
in Table \ref{Sumxc1} are all also first observations.

$\chi_{c0}$ is forbidden to decay into $K\overline{K}\pi$ 
or $\eta\pipi$ by spin-parity conservation, and only upper limits
at the 90\% C.L. are determined for these branching fractions.
For $\chi_{cJ}$ decay into hadrons in lowest-order,  
$\chi_{c1}$ decay is suppressed by a factor $\alpha_s$ compared with 
$\chi_{c2}$ decay. However, the branching fractions of $\chi_{c1}$ decays
into $K\overline{K}\pi$ and $\eta\pipi$ are both much larger than
those of $\chi_{c2}$ decays. This result needs explanation.

\acknowledgments
   The BES collaboration thanks the staff of BEPC for their 
hard efforts. This work is supported in part by the National 
Natural Science Foundation of China under contracts 
Nos. 19991480, 10225524, 10225525, the Chinese Academy
of Sciences under contract No. KJ 95T-03, the 100 Talents 
Program of CAS under Contract Nos. U-11, U-24, U-25, and 
the Knowledge Innovation Project of CAS under Contract 
Nos. U-602, U-34 (IHEP); by the National Natural Science
Foundation of China under Contract No. 10175060 (USTC); 
and by the Department of Energy under Contract 
No. DE-FG03-94ER40833 (U Hawaii).


\begin{thebibliography}{99}
\bibitem{sigma} The significance $S$ is estimated with the formula 
$S=\sqrt{-2\times(lnL_2-lnL_1)}$, where $L_1$ and $L_2$ is the 
likelihood value in the fit with and without signal, respectively.
\bibitem{evtnum}As an example, for $K^*_J(1430)^0$, the fits under different
hypotheses yield: 
$15.9\pm7.1$, $21.8\pm14.7$ and $18.6\pm11.5$ 
for $K^*_2(1430)$, $K^*_0(1430)$ and $K^*(1410)$, respectively. 
So the biggest number is
$21.8+14.7=36.5$, the smallest number is $18.6-11.5=7.1$, then the mean is
$(36.5+7.1)/2=21.8$, the error is $36.5-21.8=21.8-7.1=14.7$, and 
the average number of events is $22\pm15$.
\end{thebibliography}
\end{document}